\documentclass[]{raa}            
\usepackage{graphicx,times}
\usepackage{natbib}

\providecommand{\bm}[1]{\mbox{\boldmath$#1$\unboldmath}}

\begin{document}

   \title{Observational constraints on new generalized Chaplygin gas model
}

 \volnopage{ {\bf 0000} Vol.\ {\bf 00} No. {\bf 0}, 000--000}
   \setcounter{page}{1}

   \author{Kai Liao, Yu Pan
        \and Zong-Hong Zhu
        }

   \institute{Department of Astronomy, Beijing Normal   University, Beijing 100875, China \\
   {\it liaokai@mail.bnu.edu.cn}\\
\vs \no
   {\small Received 000; accepted 000 }
}

\abstract{We use the latest data to investigate observational constraints on
the new generalized Chaplygin gas (NGCG) model. Using the Markov Chain Monte
Carlo (MCMC) method, we constrain the NGCG model with the
type Ia supernovae (SNe Ia) from Union2 set (557 data), the usual baryonic acoustic oscillation
(BAO) observation from the spectroscopic Sloan Digital Sky Survey (SDSS) data release 7 (DR7)
galaxy sample, the cosmic microwave background (CMB) observation from the 7-year Wilkinson
Microwave Anisotropy Probe (WMAP7) results, the newly revised $H(z)$ data, as well as
a value of $\theta_{BAO} (z=0.55) = (3.90 \pm 0.38)^{\circ}$ for the
angular BAO scale. The constraint results for NGCG model are
$\omega_X$=$-1.0510_{-0.1685}^{+0.1563}(1\sigma)_{-0.2398}^{+0.2226}(2\sigma)$,
$\eta$=$1.0117_{-0.0502}^{+0.0469}(1\sigma)_{-0.0716}^{+0.0693}(2\sigma)$,
 and
$\Omega_X$=$0.7297_{-0.0276}^{+0.0229}(1\sigma)_{-0.0402}^{+0.0329}(2\sigma)$,
which give a rather stringent constraint. From the results, we can see a phantom
model is slightly favored and the probability that energy transfers from dark matter
to dark energy is a little larger than the inverse.
  }

\authorrunning{Kai Liao, Yu Pan and Zong-Hong Zhu}
\titlerunning{Observational constraints on new generalized Chaplygin gas model}
   \maketitle


%

%

\section{Introduction}
Cosmic observations suggest that the present universe is undergoing
an accelerated state (Riess et al. 1998, Perlmutter et al.  1999, Pope et al.  2004). In order
to explain this, a component with negative pressure known as dark energy was proposed.
The most simple and popular model for dark energy is cosmological constant ($\Lambda$).
This model has successfully explained many phenomena while it indeed encounters
some theoretical problems. For example, the coincidence problem and the "fine-tuning"
problem. Therefore, many other models have been proposed including quintessence (Peebles et al.  1988a, Peebles et al.  1988b),
holographic dark energy (Cohen et al.  1999, Li 2004),
quintom (Guo et al.  2005, Feng et al.  2005), phantom (Caldwell 2002), brane world (Dvali et al.  2000, Zhu et al.  2005) and so on. Among these, the generalized Chaplygin gas (GCG) model
as a unification of dark energy and dark matter is a good candidate (Bento et al.  2002a, Bento et al.  2002b).
It has being widely studied and seems to be in agreement with different observational
data (Zhu et al.  2004, Lu et al.  2009, Xu et al.  2010a, Park et al.  2010). Moreover, GCG and the original Chaplygin gas (CG) model (Kamenshchik et al.  2001)
can be connected with string and brane theory (Bilic et al.  2002).
 The equation of state (EOS) of GCG is expressed as
\begin{equation}
p_{\rm GCG} = -\frac{A}{\rho_{\rm GCG}^{\alpha}}.
\end{equation}
Under
Friedmann-Robertson-Walker (FRW) metric
with
energy momentum conservation equation , we can obtain (Amendola et al.  2003)
\begin{equation} \rho_{{\rm GCG}} =
 \rho_{{\rm GCG},0} \left[A_* +
 (1-A_*)a^{-3(1+\alpha)}\right]^{1\over(1+\alpha)},
\end{equation}
where  $A_* \equiv A/\rho_{{\rm GCG},0}^{1 + \alpha}$, $\rho_{{\rm
GCG},0}$ is the energy density today.
Besides the parameter $A$, $\alpha$ is a new parameter in contrast to the
original Chaplygin gas (CG) model. When $\alpha=1$, it becomes CG.
The core of CG and GCG is when the scale factor is very small in the early universe,
it acted as a dust, and it acts as the cosmological constant recently.
 The interacting
CG model was discussed by Ref. (Zhang. H et al.  2006) and the interacting
generalized Chaplygin gas model has been studied by Ref. (Setare et al.  2007a, Setare et al.  2007b).
Other models relevant to Chaplygin gas were proposed including modified Chaplygin gas (MCG) model (Chimento et al.  2004, Chimento et al.  2006),
variable Chaplygin gas (VCG) model (Guo et al.  2007a), extended
Chaplygin gas model (Meng et al.  2005) and new modified Chaplygin gas (NMCG) model (Chattopadhyay et al.  2008).

Since the GCG can be equal to the interacting $\Lambda$CDM model (Fabris et al.  2004, Zhang. X et al.  2006),
a new generalized Chaplygin gas (NGCG) model which equals to a kind of interacting XCDM model was proposed as a unification
of X-type dark energy with EOS parameter $\omega_X$ and cold dark matter (Zhang. X et al.  2006).
The interacting new generalized Chaplygin gas model has been discussed by Ref. (Jamil et al.  2010).

There are different kinds of observational data which can be used to constrain cosmological models: type supernovae Ia,
observational Hubble parameter data, baryon acoustic oscillation (BAO), cosmic microwave background
data, lensing (Liao et al.  2012a) and so on. For BAO, using spectroscopic determinations
of the redshifts of the galaxies, several detections at different redshifts have been studied (Percival et al.  2007, Percival et al.  2010). This
method studies the three-dimensional averaged distance parameter $D_V$. The radial BAO Data was discussed by Ref. (Gaztanaga et al.  2009).
Recently, a new determination of the BAO scale using the photometric sample of Luminous Red Galaxies in DR7
is performed (Crocce et al.  2011, Carnero et al.  2011). They get a value of $\theta_{BAO} (z=0.55) = (3.90 \pm 0.38)^{\circ}$, including systematic errors, for the
angular BAO scale. It is the first direct measurement of the pure angular BAO signal. Combined with previous
BAO signals, it can break the degeneracies in the determination of model parameters.

In this paper, we use the Markov chain Monte Carlo (MCMC) method to constrain NGCG model
from the latest data including the BAO data at $z=0.55$. In Section 2, we give a brief introduction of the NGCG model. In Section 3,
we introduce the observational data. The full parameter space using different combinations of data
is provided in Section 4. At last, we summarize our results in Section 5.

\section{The new generalized Chaplygin gas model} \label{sec2}
We give a very brief introduction to the new generalized Chaplygin gas model in this section. For details of this model, please refer to Ref. (Zhang. X et al.  2006).
Assuming the universe is flat with the FRW metric, the equation of state of NGCG fluid (Zhang. X et al.  2006) is
\begin{equation}
p_{\rm NGCG} = -\frac{\tilde{A}(a)}{\rho_{\rm NGCG}^{\alpha}},
\end{equation}
where $a$ is the scale factor and $\alpha$ is a parameter. NGCG fluid
consists of dark energy $\rho_X\sim a^{-3(1+w_X)}$, where $w_X$ is the
EOS parameter, and dark matter $\rho_{DM}\sim a^{-3}$. Naturally, the
energy density of the NGCG can be considered as
\begin{equation}
\rho_{\rm NGCG}=  \left[A a^{-3(1+w_X)(1+\alpha)} + {B a^{-3 (1 +
\alpha)}}\right]^{1 \over 1 + \alpha}~,
\end{equation}
with
\begin{equation}
A+B=\rho_{\rm NGCG,0}^\eta~,
\end{equation}
where $A$ and $B$ positive constants. Considering
the energy conservation equation, $\tilde{A}(a)$ can be acquired as
\begin{equation}
\tilde{A}(a)=-w_XAa^{-3(1+w_X)(1+\alpha)}~.
\end{equation}
We can easily see it becomes GCG
when the EOS parameter of dark energy component $w_X$ is $-1$. On the
other hand it becomes XCDM if $\alpha=0$. It is remarkable that $\alpha$
describes the interacting between dark energy and dark matter. When
$\alpha>0$, the energy transfers from dark matter to dark energy. On
the contrary, the energy transfers from dark energy to dark matter in
the case $\alpha<0$. Based on Ref. (Zhang. X et al.  2006), we take the radiation
component into consideration, which dominated the early universe. The
whole density consists of NGCG component, the baryon matter component
and the radiation component $\rho_{tot}=\rho_{NGCG}+\rho_{b}+\rho_{r}$.
The Friedmann equation can be expressed as
\begin{equation}
H(a)=H_0E(a),
\end{equation}
where
\begin{eqnarray}
E(a)^2 &=& (1-\Omega_b-\Omega_r)a^{-3}\left[1-{\Omega_X\over
1-\Omega_b-\Omega_r}(1-a^{-3w_X\eta})\right]^{1/\eta}\nonumber\\&&
+\Omega_ba^{-3}+\Omega_ra^{-4}~.
\end{eqnarray}
The $H_0$, $\Omega_X$, $\Omega_b$ and $\Omega_r$ represent Hubble
constant, dimensionless dark energy, baryon matter and radiation today.
Consist with the original paper, the parameter $\eta=1+\alpha$.

\section{Current observational data}\label{sec3}
Now, we introduce the methods of constraints on NGCG model
by using the latest data.

\subsection{Baryon acoustic oscillation including data at $z=0.55$}
For BAO, the distance scale is defined as (Eisenstein et al.  2005)
\begin{equation} D_V(z)=\frac{1}{H_0}\big
[\frac{z}{E(z)}\big(\int_0^{z}\frac{dz}{E(z)}\big
)^2\big]^{1/3}~,
\end{equation}
and baryons were released from photons at the so-called drag epoch, the corresponding redshift $z_d$ is
give by
\begin{equation}
z_{d}=\frac{1291(\Omega_{\mathrm{m0}}h^2)^{0.251}}{[1+0.659(\Omega_\mathrm{m0}h^2)^{0.828}]}[(1+b_{1}(\Omega_{b}h^2)^{b_2})],
\end{equation}
where
$b_1=0.313(\Omega_{\mathrm{m0}}h^2)^{-0.419}[1+0.607(\Omega_{\mathrm{m0}}h^2)^{0.674}]^{-1}$
and $b_2=0.238(\Omega_{\mathrm{m0}}h^2)^{0.223}$ (Eisenstein et al.  1998).
For observation of BAO, we choose the measurement of the distance radio ($d_z$) at $z=0.2$ and $z=0.35$ (Percival et al.  2010).
The definition is given by
\begin{equation}
d_z=\frac{r_s(z_d)}{D_V(z)},
\end{equation}
where $r_s(z_d)$ is the comoving sound horizon.
The SDSS data release 7 (DR7) galaxy sample gives the best-fit values
of the data set ($d_{0.2}$, $d_{0.35}$) (Percival 2010)
\begin{eqnarray}
\hspace{-.5cm}\bar{\bf{P}}_{matrix} &=& \left(\begin{array}{c}
{\bar d_{0.2}} \\
{\bar d_{0.35}}\\
\end{array}
  \right)=
  \left(\begin{array}{c}
0.1905\pm0.0061\\
0.1097\pm0.0036\\
\end{array}
  \right).
 \end{eqnarray}
The $\chi^2$ value of this BAO observation from SDSS DR7 can be
calculated as (Percival et al.  2010)
\begin{eqnarray}
\chi^2_{matrix}=\Delta
\textbf{P}_{matrix}^\mathrm{T}{\bf
C_{matrix}}^{-1}\Delta\textbf{P}_{matrix},
\end{eqnarray}
where  the corresponding inverse  covariance matrix is
\begin{eqnarray}
\hspace{-.5cm} {\bf C_{matrix}}^{-1}=\left(
\begin{array}{ccc}
30124& -17227\\
-17227& 86977\\
\end{array}
\right).
\end{eqnarray}

Moreover, a new determination of BAO scale using photometric sample of Luminous
Red Galaxies in the DR7 is presented (Crocce et al.  2011, Carnero et al.  2011). It make use of $\sim 1.5 \times 10^6$ luminous red galaxies
with photometric redshifts. They get a value of $\theta_{BAO} (z=0.55) = (3.90 \pm 0.38)^{\circ}$ for the
angular BAO scale including systematic errors. It is the first direct measurement of the pure angular BAO signal.
The definition of $\theta_{BAO}$ is expressed as (Sanchez et al.  2010)
\begin{equation}
\theta_{BAO}=\frac{r_s(z_d)}{\chi(z)},
\end{equation}
where the comoving radial distance $\chi(z)$ depending on NGCG model is defined as
\begin{equation}
\chi(z) = \frac{1}{H_0} \int_{0}^{z}\frac{dz}{E(a)} \,\, .
\label{comovingRadial}
\end{equation}
The corresponding $\chi^2$ of this BAO data is expressed as

\begin{equation}
\chi^2_{angular}=\frac{[\theta_{BAO}-0.0681]^2}{0.00663^2},
\end{equation}
where the unit has been transformed into radian.
Since this data is independent of previous signals,
the total $\chi^2$ can be acquired as
\begin{equation}
\chi^2_{BAO}=\chi^2_{matrix}+\chi^2_{angular}.
\end{equation}

\subsection{Observational Hubble parameter data}
It is known that we use the distance scale of SNe Ia, CMB and BAO to constrain
cosmological parameters. However, the distance scale is determined by integrating
the Hubble parameter, which can not reflect the fine structure of $H(z)$. Thus,
investigating the $H(z)$ data directly reveals a more real scene. Many works have
done using the newly revised $H(z)$ (Xu et al.  2010b, Ma et al.  2011).
The measurement of Hubble parameter data as a function of redshift $z$
depends on the differential ages of red-envelope galaxies
\begin{equation}
H(z)=-\frac{1}{1+z}\frac{dz}{dt}\,.
\end{equation}
We choose 12 data from this method (Riess et al.  2009, Stern et al.  2009). In addition,
the data can be obtained from the BAO scale as a standard ruler in
the radial direction. So, we choose another three data at different
redshifts (Gaztanaga et al.  2008).

The $\chi^2$ value of the $H(z)$ data can be expressed as
\begin{equation}
\label{chi2H}
\chi^2_H=\sum_{i=1}^{15}\frac{[H(z_i)-H_{obs}(z_i)]^2}{\sigma_{i}^2},
\end{equation}
where $\sigma_{i}$ is the $1\sigma$ uncertainty of the observational
$H(z)$ data.

\subsection{Type Ia supernovae}
SNe Ia has been playing an important role in studying cosmology
since it first revealed the accelerated expansion of the universe.
The recent data (Union2) is given by the Supernova Cosmology Project (SCP)
collaboration including 557 samples (Amanullah et al.  2010).
The new data has been used to constrain cosmological models (Xu et al.  2010c, Xu et al.  2010d, Liao et al.  2012b). In practice, people
use the distance modules of supernovae to reflect the cosmological model and
constrain the cosmological parameters. The
distance modules is determined by the luminosity distance
\begin{equation}
\mu=5 \log(d_L/\rm{Mpc})+25~,
\end{equation}
where $d_L$ is the luminosity distance. In the flat universe, it is connected
with redshift which is a observational quantity
\begin{equation}
d_L={(1+z)} \int^{z}_0{dz'}/{H(z')}.
\end{equation}
We choose the marginalized nuisance parameter (Nesseris et al.  2005) for
$\chi^2$:
\begin{equation}
\label{chi2SN} \chi^2_{\rm
SNe}=A-\frac{B^2}{C},
\end{equation}
where $A=\sum_i^{557}{(\mu^{\rm data}-\mu^{\rm
theory})^2}/{\sigma^2_{i}}$, $B=\sum_i^{557}{(\mu^{\rm
data}-\mu^{\rm theory})}/{\sigma^2_{i}}$, $C=\sum_i^{557}{1}/{\sigma^2_{i}}$,
$\sigma_{i}$ is the 1$\sigma$ uncertainty of the observational data.

\subsection{Cosmic microwave background}
For CMB, the acoustic scale is related to the distance ratio and is
expressed as
\begin{equation}
l_a=\pi\frac{\Omega_\mathrm{k}^{-1/2}sinn[\Omega_\mathrm{k}^{1/2}\int_0^{z_{\ast}}\frac{dz}{E(z)}]/H_0}{r_s(z_{\ast})},
\end{equation}
where $r_s(z_{\ast})
={H_0}^{-1}\int_{z_{\ast}}^{\infty}c_s(z)/E(z)dz$ is the comoving
sound horizon at photo-decoupling epoch. The CMB shift parameter R
is expressed as (Bond et al.  1997)
\begin{equation}
R=\Omega_{\mathrm{m0}}^{1/2}\Omega_\mathrm{k}^{-1/2}sinn\bigg[\Omega_\mathrm{k}^{1/2}\int_0^{z_{\ast}}\frac{dz}{E(z)}\bigg],
\end{equation}
where the redshift $z_{\ast}$ corresponding to the decoupling epoch of photons is given by (Hu et al.  1996)
\begin{equation}
z_{\ast}=1048[1+0.00124(\Omega_bh^2)^{-0.738}(1+g_{1}(\Omega_{\mathrm{m0}}h^2)^{g_2})],
\end{equation}
where
$g_1=0.0783(\Omega_bh^2)^{-0.238}(1+39.5(\Omega_bh^2)^{-0.763})^{-1}$,
$g_2=0.560(1+21.1(\Omega_bh^2)^{1.81})^{-1}$.

For the CMB observation, we choose the data set including the the acoustic
scale ($l_a$), the shift parameter ($R$), and the redshift of
recombination ($z_{\ast}$). The WMAP7 measurement gives the best-fit
values of the data set (Komatsu et al.  2011)
\begin{eqnarray}
\hspace{-.5cm}\bar{\textbf{P}}_{\rm{CMB}} &=& \left(\begin{array}{c}
{\bar l_a} \\
{\bar R}\\
{\bar z_{\ast}}\end{array}
  \right)=
  \left(\begin{array}{c}
302.09 \pm 0.76\\
1.725\pm 0.018\\
1091.3 \pm 0.91 \end{array}
  \right).
 \end{eqnarray}
 The $\chi^2$ value of the CMB observation can be calculated as
(Komatsu et al.  2011)
\begin{eqnarray}
\chi^2_{\mathrm{CMB}}=\Delta
\textbf{P}_{\mathrm{CMB}}^\mathrm{T}{\bf
C_{\mathrm{CMB}}}^{-1}\Delta\textbf{P}_{\mathrm{CMB}},
\end{eqnarray}
where $\Delta\bf{P_{\mathrm{CMB}}} =
\bf{P_{\mathrm{CMB}}}-\bf{\bar{P}_{\mathrm{CMB}}}$, and the
corresponding inverse  covariance matrix is
\begin{eqnarray}
\hspace{-.5cm} {\bf C_{\mathrm{CMB}}}^{-1}=\left(
\begin{array}{ccc}
2.305 &29.698 &-1.333\\
29.698 &6825.270 &-113.180\\
-1.333 &-113.180 &3.414
\end{array}
\right).
\end{eqnarray}

\begin{table*}
\tiny
 \begin{center}
 \begin{tabular}{|c|c|c|c|} \hline\hline
 & \multicolumn{3}{c|}{The NGCG Model}  \\
 \cline{2-4}                 &     SNe+CMB+BAO+$H(z)$                  &     SNe+BAO+CMB                       &     CMB+BAO+$H(z)$          \\ \hline
$\Omega_{\rm{b}}h^2$  \ \ & \ \ $0.0224_{-0.0010}^{+0.0012}(1\sigma)_{-0.0014}^{+0.0017}(2\sigma)$ \ \  & \ \ $0.0224_{-0.0011}^{+0.0012}(1\sigma)_{-0.0015}^{+0.0017}(2\sigma)$ \ \ & \ \ $0.0225_{-0.0011}^{+0.0011}(1\sigma)_{-0.0015}^{+0.0016}(2\sigma)$\ \ \\
$\Omega_{\rm{DM}}h^2$  \ \ & \ \ $0.1139_{-0.0075}^{+0.0079}(1\sigma)_{-0.0102}^{+0.0109}(2\sigma)$ \ \  & \ \ $0.1144_{-0.0087}^{+0.0072}(1\sigma)_{-0.0118}^{+0.0106}(2\sigma)$ \ \ & \ \ $0.1153_{-0.0097}^{+0.0084}(1\sigma)_{-0.0138}^{+0.0113}(2\sigma)$\ \ \\
$\omega_X$                 \ \ & \ \ $-1.0510_{-0.1685}^{+0.1563}(1\sigma)_{-0.2398}^{+0.2226}(2\sigma)$ \ \  & \ \ $-1.0297_{-0.1829}^{+0.1765}(1\sigma)_{-0.2640}^{+0.2474}(2\sigma)$ \ \ & \ \ $-1.1156_{-0.3150}^{+0.2786}(1\sigma)_{-0.4495}^{+0.3588}(2\sigma)$\ \ \\
$\eta$              \ \ & \ \ $1.0117_{-0.0502}^{+0.0469}(1\sigma)_{-0.0716}^{+0.0693}(2\sigma)$ \ \ & \ \ $1.0076_{-0.0504}^{+0.0502}(1\sigma)_{-0.0759}^{+0.0731}(2\sigma)$\ \  & \ \ $1.0170_{-0.0598}^{+0.0477}(1\sigma)_{-0.0907}^{+0.0673}(2\sigma)$\ \ \\
$\Omega_X$    \ \ & \ \ $0.7297_{-0.0276}^{+0.0229}(1\sigma)_{-0.0402}^{+0.0329}(2\sigma)$\ \   & \ \ $0.7232_{-0.0298}^{+0.0305}(1\sigma)_{-0.0446}^{+0.0408}(2\sigma)$\ \  & \ \ $0.7367_{-0.0382}^{+0.0351}(1\sigma)_{-0.0543}^{+0.0458}(2\sigma)$\ \ \\
Age/Gyr               \ \ & \ \ $13.73_{-0.12}^{+0.14}(1\sigma)_{-0.18}^{+0.20}(2\sigma)$\ \            & \ \ $13.75_{-0.14}^{+0.16}(1\sigma)     _{-0.20}^{+0.24}(2\sigma)$\ \      & \ \ $13.70_{-0.14} ^{+0.21}(1\sigma)    _{-0.19}^{+0.33}(2\sigma)$\ \ \\
$\Omega_{\rm{m}}$     \ \ & \ \ $0.2703_{-0.0229}^{+0.0276}(1\sigma)_{-0.0329}^{+0.0402}(2\sigma)$\ \   & \ \ $0.2768_{-0.0305}^{+0.0298}(1\sigma)  _{-0.0408}^{+0.0446}(2\sigma)$\ \    & \ \ $0.2633_{-0.0351}^{+0.0382}(1\sigma)_{-0.0458}^{+0.0543}(2\sigma)$\ \ \\
$H_0$                 \ \ & \ \ $71.03_{-3.37}^{+3.20}(1\sigma)_{-4.96}^{+4.55}(2\sigma)$\ \            & \ \ $70.29_{-3.76}^{+3.86}(1\sigma)     _{-5.34}^{+5.61}(2\sigma)$\ \      & \ \ $72.36_{-5.89}  ^{+6.16}(1\sigma) _{-7.67}  ^{+8.45}(2\sigma)$\ \ \\
 \hline\hline
$\chi_{\rm min}^2$              \ \ & \ \ $543.668$\ \                  & \ \ $533.586$\ \                   & \ \ $ 12.701$\ \      \\
\hline\hline

 \end{tabular}
 \end{center}\label{tab:resultsNGCG}
 \small{\caption{The best-fit values of  parameters $\Omega_{\rm{b}}h^2$, $\Omega_{\rm{DM}}h^2$ $\omega_X$, $\eta$,
 $\Omega_X$, Age/Gyr, $\Omega_{m}$, and $H_0$ for NGCG model
 with the 1-$\sigma$ and 2-$\sigma$ uncertainties, as well as $\chi_{\rm min}^2$,
for the data sets  SNe+CMB+BAO+$H(z)$, SNe+BAO+CMB, and CMB+BAO+$H(z)$,
respectively. }}\label{tab:resultsNGCG}
 \end{table*}

\begin{figure}
\includegraphics[width=15cm]{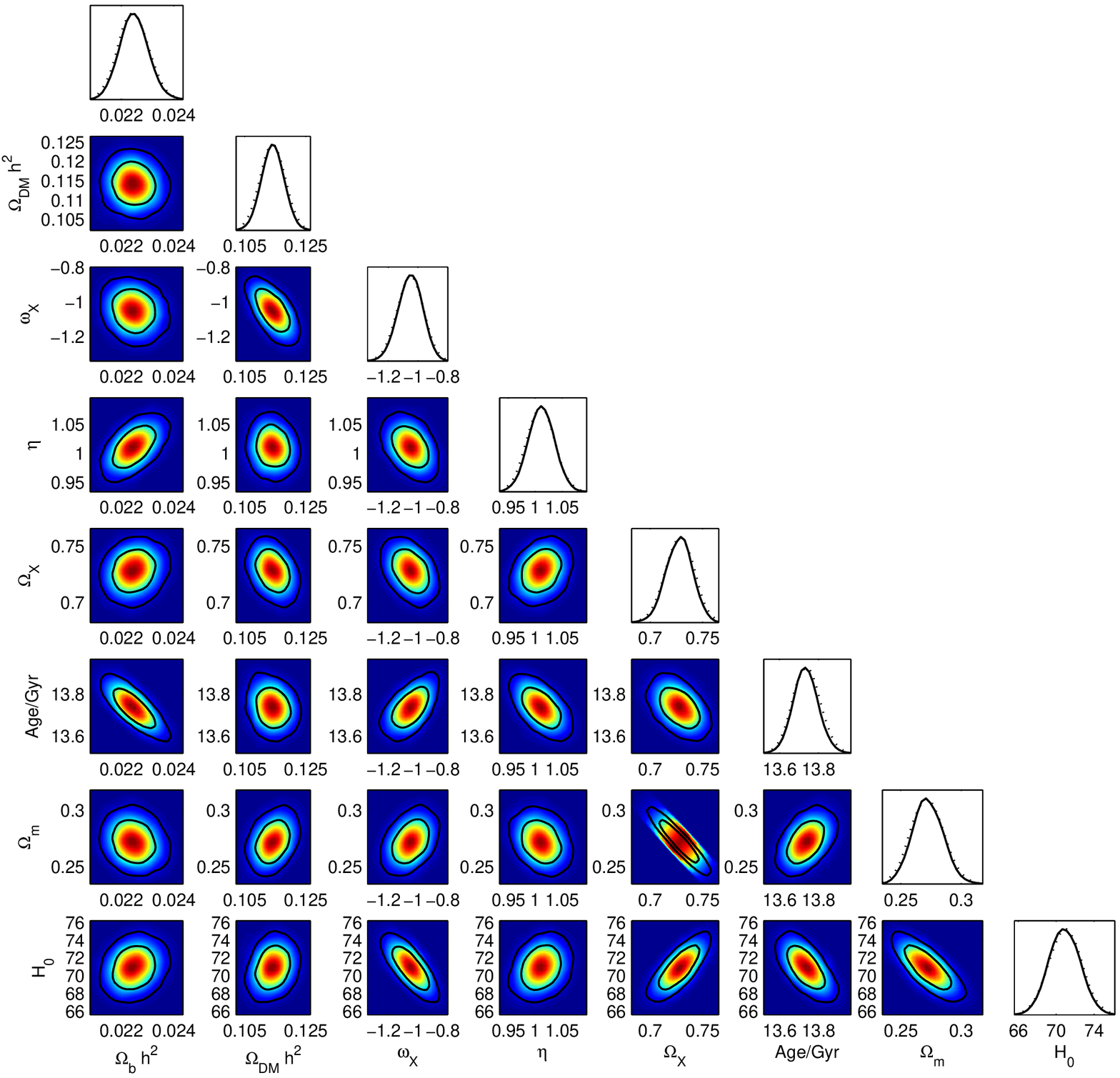}
\caption{The 2-D regions and 1-D marginalized distribution with the
1-$\sigma$ and 2-$\sigma$ contours of parameters
$\Omega_{\rm{b}}h^2$, $\Omega_{\rm{DM}}h^2$ $\omega_X$, $\eta$,
 $\Omega_X$, Age/Gyr, $\Omega_{m}$, and $H_0$  in NGCG model, for the data sets
SNe+CMB+BAO+$H(z)$.
\label{Fig1}}
\end{figure}

\begin{figure}
\includegraphics[width=15cm]{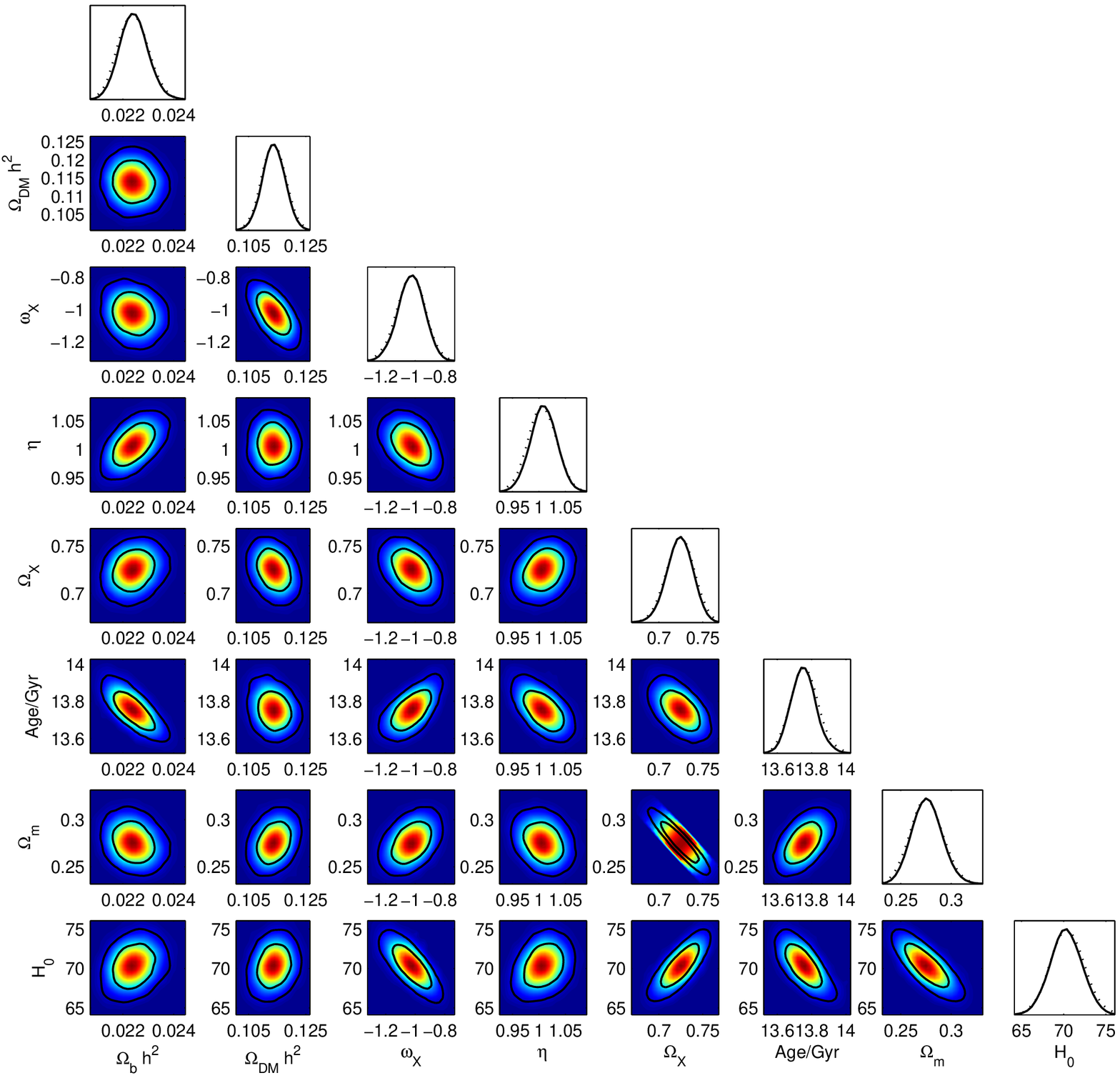}
\caption{The 2-D regions and 1-D marginalized distribution with the
1-$\sigma$ and 2-$\sigma$ contours of parameters
$\Omega_{\rm{b}}h^2$, $\Omega_{\rm{DM}}h^2$ $\omega_X$, $\eta$,
 $\Omega_X$, Age/Gyr, $\Omega_{m}$, and $H_0$  in NGCG model, for the data sets
SNe+CMB+BAO.
\label{Fig2}}
\end{figure}

\begin{figure}
\includegraphics[width=15cm]{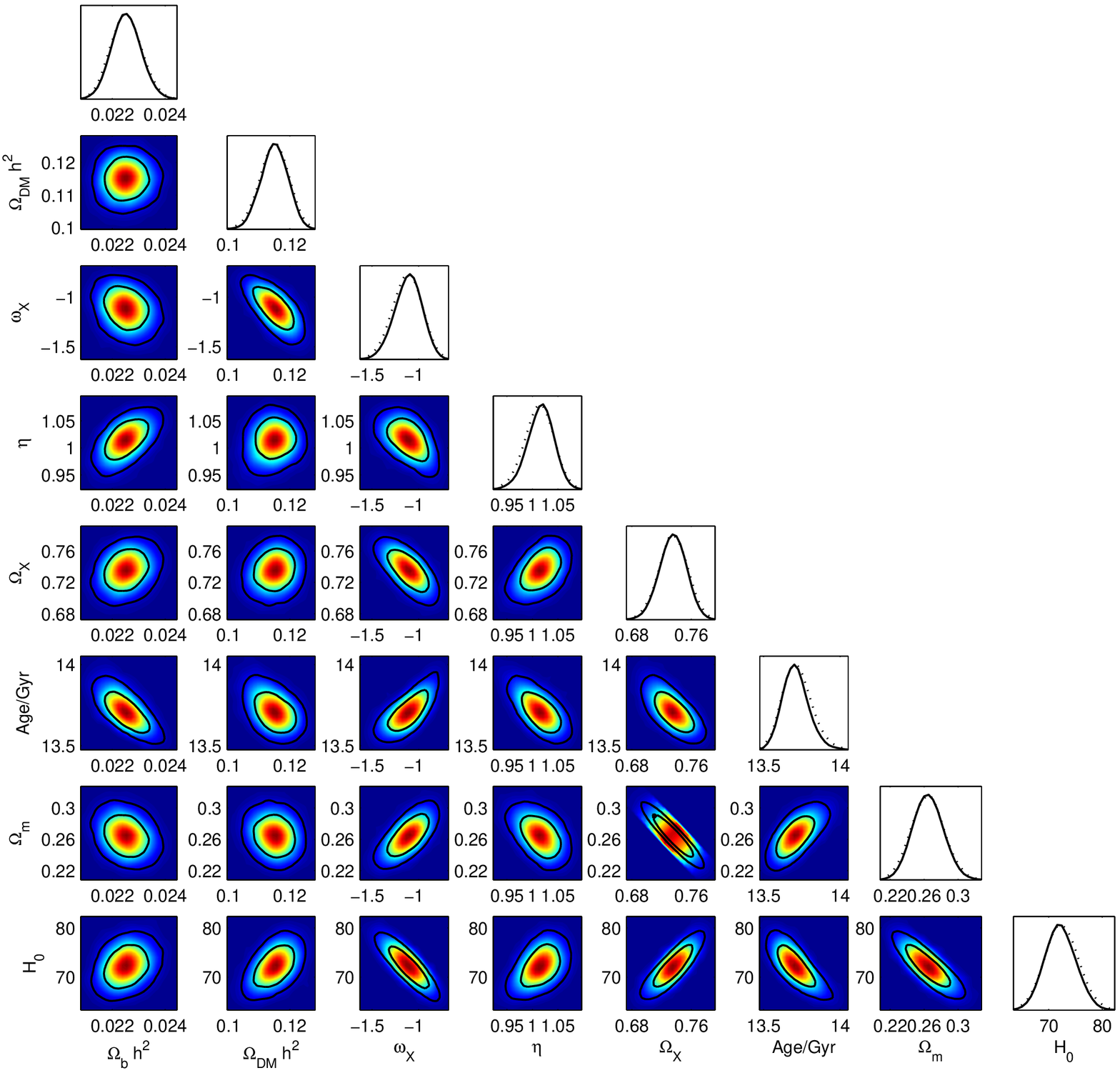}
\caption{The 2-D regions and 1-D marginalized distribution with the
1-$\sigma$ and 2-$\sigma$ contours of parameters
$\Omega_{\rm{b}}h^2$, $\Omega_{\rm{DM}}h^2$ $\omega_X$, $\eta$,
 $\Omega_X$, Age/Gyr, $\Omega_{m}$, and $H_0$  in NGCG model, for the data sets
CMB+BAO+$H(z)$.
\label{Fig3}}
\end{figure}

\section{Constraints on NGCG model via MCMC method}\label{sec4}
In order to constrain the parameters of NGCG model, We take the usual
maximum likelihood method of $\chi^{2}$ fitting with the Markov Chain
Monte Carlo (MCMC) method. We take the Metropolis-Hastings algorithm
with uniform prior probability distribution. By using Monto Carlo method,
we generate a chain of sample points distributed in the parameter space
according to the posterior probability, then repeat the process until the
established convergence accuracy is acquired. In our testing, the convergence
of the chains $R-1$ is set to be less than 0.003 which is small enough to
guarantee the accuracy. The code we use is based on CosmoMCMC (Lewis et al.  2002).
We combine the SNe, BAO, CMB and $H(z)$ data by multiplying the likelihood functions
to constrain the NGCG model. The total $\chi^{2}$ can be expressed as
\begin{equation}
\chi^2=\chi^2_{\rm SNe}+\chi^2_{\rm BAO}+\chi^2_{\rm CMB}+\chi^2_{H}.
\end{equation}

We show the 1-D probability of each parameter ($\Omega_{\rm{b}}h^2$, $\Omega_{\rm{dm}}h^2$
$\omega_X$, $\eta$, $\Omega_X$, Age/Gyr, $\Omega_{\rm{m}}$, $H_0$)
($\Omega_{\rm{b}}$, $\Omega_{\rm{dm}}$, $\Omega_X$ and $\Omega_{\rm{m}}$
are responding to baryon matter, dark matter, dark energy and the total matter)
and 2-D plots for parameters between each other for the NGCG model with
SNe + CMB + BAO + $H(z)$ in Fig. \ref{Fig1}. The best-fit values of NGCG model parameters
with the four kinds of data above are
$\omega_X$=$-1.0510_{-0.1685}^{+0.1563}(1\sigma)_{-0.2398}^{+0.2226}(2\sigma)$,
$\eta$=$1.0117_{-0.0502}^{+0.0469}(1\sigma)_{-0.0716}^{+0.0693}(2\sigma)$,
$\Omega_bh^2$=$0.0224_{-0.0010}^{+0.0012}(1\sigma)_{-0.0014}^{+0.0017}(2\sigma)$,
$\Omega_X$=$0.7297_{-0.0276}^{+0.0229}(1\sigma)_{-0.0402}^{+0.0329}(2\sigma)$,
and the $\chi_{\rm min}^2=543.668$.
We also constrain NGCG model with other combinations of data for comparison: SNe + CMB + BAO
and CMB + BAO + $H(z)$. They are showed in Fig. \ref{Fig2} and Fig. \ref{Fig3}. The best-fit values
of each parameter with the 1-$\sigma$ and 2-$\sigma$ uncertainties and the $\chi_{\rm min}^2$
are presented in Table 1.
We can see the combination of the latest data gives a rather tight constraint on NGCG parameters
especially for $\eta$ which describes the interacting between dark energy and dark matter.
The best-fit of $\omega_X$ is slightly smaller than -1, which makes a little difference
from Zhang et al. 2006, and $\eta$ is a little larger than 1. Considering the uncertainty,
these indicate the scene of the universe that the NGCG model reflects is near the $\Lambda$CDM model.
On the other hand, the results still accommodate a interacting XCDM model. In 1-$\sigma$ range for
SNe + CMB + BAO + $H(z)$, $\omega_X=(-1.2195, -0.8947)$, $\alpha=\eta-1=(-0.0385, 0.0585)$.
The result of $\omega_X$
indicates the dark energy acts as a phantom is slightly favored. The constraint on $\alpha$ indicates the probability that
energy transfers from dark matter to dark energy is a little larger than the inverse. Moreover, since $\alpha$ is constrained to a very small range
near 0, the interacting between dark sectors seems very weak.
The results also totally rule out the original
CG model and the VCG model which require $\alpha=1$. This is in agreement with (Zhang. X et al.  2006, Wu et al.  2007).

\section{Conclusion}\label{sec5}
With the MCMC method, we give constraints on the NGCG model from the
latest data including the SNe Ia data from Union2, BAO data from
SDSS DR7 galaxy sample, and CMB observation from WMAP7, as well as the newly obtained
angular BAO signal $\theta_{BAO} (z=0.55) = (3.90 \pm 0.38)^{\circ}$,
which is the first direct measurement of the pure angular BAO signal. The best-fit values
of the parameters for NGCG are
$\omega_X$=$-1.0510_{-0.1685}^{+0.1563}(1\sigma)_{-0.2398}^{+0.2226}(2\sigma)$,
$\eta$=$1.0117_{-0.0502}^{+0.0469}(1\sigma)_{-0.0716}^{+0.0693}(2\sigma)$,
$\Omega_bh^2$=$0.0224_{-0.0010}^{+0.0012}(1\sigma)_{-0.0014}^{+0.0017}(2\sigma)$,
$\Omega_X$=$0.7297_{-0.0276}^{+0.0229}(1\sigma)_{-0.0402}^{+0.0329}(2\sigma)$,
and the $\chi_{\rm min}^2=543.668$,
which give a quite tight constraint. From the results we can see
$\Lambda$CDM model is near the best-fit point, it remains a good choice
for explaining the observation. However, the NGCG model permits a interacting
XCDM model. The constraint results seem to slightly favor a phantom and the energy transfers from
dark matter to dark energy. Moreover, since $\alpha$ is constrained to a very small range
near 0, the interacting between dark sectors seems very weak. The results are consist with the
situations in Ref. (Guo 2007b et al. , Chen 2010 et al. , Feng 2008 et al. , Cao 2011a et al. , Cao 2011b et al. ). These papers as well as ours manifest current observations can not
distinguish the directions of energy transforming well, thus the coincidence problem can not be solved so far in this way. For future study on coincidence problem, we hope more data and other independent observations can
give the distinguishability. Besides, since the constraint results for these interacting XCDM models slightly favor the energy transfers
from dark matter to dark energy, we may suspect whether this way is valid for explaining the coincidence problem.

\begin{acknowledgements}
We are grateful to Lixin Xu  for introducing the MCMC method.
We also thank Shuo Cao and Hao Wang for helpful discussions.
This work was supported by the National Natural Science Foundation of
China under the Distinguished Young Scholar Grant 10825313,
the Ministry of Science and Technology national basic science Program (Project 973)
under Grant No.2012CB821804, the Fundamental Research Funds for the Central
Universities and Scientific Research Foundation of Beijing Normal University.
\end{acknowledgements}


\end{document}